\documentclass[letterpaper,11pt]{article}

\usepackage{amssymb}
\usepackage{amsmath}
\usepackage{cite}
\usepackage{graphicx}
\usepackage{epsfig}
\usepackage{epstopdf}

\numberwithin{equation}{section}

\newcommand {\beq} {\begin{equation}}
\newcommand {\eeq} {\end{equation}}
\newcommand {\beqa} {\begin{eqnarray}}
\newcommand {\eeqa} {\end{eqnarray}}
\newcommand {\beqan} {\begin{eqnarray*}}
\newcommand {\eeqan} {\end{eqnarray*}}

\newcommand {\ph}[1]{\phantom{#1}}
\newcommand {\ie}{i.e.~}
\newcommand {\eg}{e.g.~}

\newcommand {\sss} {\scriptscriptstyle}

\newcommand{\Q}{\ensuremath{\mathcal{Q}}}

\newcommand{\pbf}{\ensuremath{\mathbf{p}}}

\newcommand{\kbf}{\ensuremath{\mathbf{k}}}

\newcommand{\al}{\ensuremath{\alpha}}

\newcommand{\ga}{\ensuremath{\gamma}}

\newcommand{\de}{\ensuremath{\delta}}

\newcommand{\eps}{\ensuremath{\epsilon}}

\newcommand{\si}{\ensuremath{\sigma}}

\newcommand{\om}{\ensuremath{\omega}}

\newcommand{\mathHb}[1]{{\mathop{\kern0pt#1}\limits^{\,\sss
      \prime\prime}\vphantom{#1}}}

\newcommand {\pa} {\partial}

\newcommand{\eqnlab}[1]{\label{eqn:#1}}
\newcommand{\figlab}[1]{\label{fig:#1}}

\newcommand{\eqnref}[1]{(\ref{eqn:#1})}
\newcommand{\figref}[1]{\ref{fig:#1}}

\newcommand{\Eqnref}[1]{Eq.~(\ref{eqn:#1})}

\newcommand{\Eqsref}[1]{Eqs.~(\ref{eqn:#1})}

\newcommand{\seclab}[1]{\label{sec:#1}}

\newcommand{\secref}[1]{\ref{sec:#1}}

\begin{document}

 \pagestyle{empty}
 \vskip-10pt

\begin{center}

\vspace*{2cm}

\noindent
{\LARGE\textsf{\textbf{Scattering in $D=5$ super Yang-Mills theory \\[5mm] and the relation to (2,0) theory}}}
\vskip 1truecm
\vskip 2truecm

{\large \textsf{\textbf{Erik Flink}}} \\ 
\vskip 1truecm
{\tt erik.flink@chalmers.se}
\vskip 0.5truecm
{\it Department of Fundamental Physics\\ Chalmers University of
  Technology\\ SE-412 96 G\"{o}teborg,
  Sweden}\\
\end{center}
\vskip 1cm
\noindent{\bf Abstract:}
Compactifying the $A_1$ version of (2,0) theory on a circle gives rise to five-dimensional, maximally supersymmetric Yang-Mills theory. In the Coulomb branch, where the $SU(2)$ gauge group is spontaneously broken to a $U(1)$ subgroup, the degrees of freedom are constituted by one massless and two massive vector multiplets. Because of the relation to the six-dimensional (2,0) theory, we are then interested in scattering processes where both the in-state and the out-state consist of one massless and one massive particle. We show that the corresponding part of the $S$ matrix is determined by the symmetries of the theory up to a single unknown function, which depends on the energy and mass of the incoming particles, together with the scattering angle. Performing a straight forward scattering calculation by means of Feynman diagrams, this function is determined to leading order in a low-energy approximation. The result is strikingly simple, and it coincides exactly with the corresponding function in the (2,0) theory.

\newpage
\pagestyle{plain}

\tableofcontents

\section{Introduction}
\subsection{Motivation}
In a series of papers we have investigated the so called $A_1$ version of the six-dimensional (2,0) theories. This is the theory describing two parallel and nearby $M5$-branes with membranes stretching between them. These membranes are perceived as strings on the world-volume of the $M5$-branes \cite{Strominger:1996,Howe:1998}. The fluctuations of the $M5$-branes are described by a tensor multiplet living on the branes \cite{Kaplan:1995,Dasgupta:1995,Witten:1995b}. In this $A_1$ version there exist exactly one tensor multiplet and only one type of string.

When compactifying this theory on a circle one obtains maximally supersymmetric Yang-Mills theory with $SU(2)$ gauge group spontaneously broken to a $U(1)$ subgroup \cite{Witten:1995} (see \cite{Seiberg:1997} for a short and nice review). The degrees of freedom are constituted by one massless neutral vector multiplet and two oppositely charged, massive vector multiplets. The relation between the degrees of freedom in the five- and six-dimensional theories was thoroughly established in \cite{Gustavsson:2001b}.

 In \cite{Flink:2005} we studied the scattering of a single tensor multiplet particle against an infinitely long tensile string. In five dimensions, this type of process corresponds to the scattering of a massless vector multiplet particle against a massive vector multiplet particle. We found that in the six-dimensional problem, symmetries provided much information about the $S$ matrix; in fact it was determined by the symmetries up to a single unknown function of the particle's energy, the string tension and the scattering angles. We calculated this function using a specific model of the theory, proposed in \cite{AFH:2003sc}, and the result was strikingly simple. The high amount of symmetry combined with the simplicity of the calculated cross-sections made it seem very interesting to examine whether the analogous five-dimensional problem shows any similar behaviour. We will find that it actually does: Just as in six dimensions, only one single function needs to be determined by an actual scattering calculation. Furthermore, we show that to first order in a low-energy expansion, this function is equal to its six-dimensional counterpart.  We then argue that this may be interpreted as a confirmation of the belief that (2,0) theory should provide the UV-completion of five-dimensional super Yang-Mills theory.

\subsection{Preliminaries}
\seclab{preliminaries}
We start by noticing that there are sixteen supercharges in five-dimensional maximally supersymmetric Yang-Mills theory. Under the symmetry group $SO(1,4) \times SO(5)_R$ they transform as $(4;4)$ and they are subject to a symplectic Majorana reality constraint. However, in the scattering processes of our concern, the existence of massive particles will break the $R$-symmetry to $SO(4)_R$ (we explain this in the beginning of section \secref{massive vm}). Furthermore, we will take the incoming massless particle to have spatial momentum $\pbf$ pointing in the $x^4$-direction. The incoming massive particle will have spatial momentum $\kbf = \bf 0$. Finally, we will call the spatial momenta of the outgoing particles $\pbf'$ and $\kbf'$. By conservation of momentum $\pbf$, $\pbf'$ and $\kbf'$ must all lie in the same plane, which we can take to be the $x^3x^4$-plane without loss of generality. This setup breaks the Lorentz group to an $SO(2)$ subgroup of rotations in the $x^1x^2$-plane. We call the generator of rotations in that plane $\hat{J}$, which is normalized so that $\exp{i\al \hat{J}}$ is a rotation by an angle $\al$. (In the spinor representation this means that $\hat{J}^2= 1/4$.) Hence, the preserved symmetry group is $SO(2)_{\pbf,\kbf,\pbf',\kbf'} \times SO(4)_R$ and it is convenient to label the states of the massless and massive vector multiplets by giving their $SO(4)_R$ representations together with the eigenvalue under $\hat{J}$ as a subscript.

We end this short section by giving the Clifford algebras for the $SO(1,4)$ Lorentz group and the $SO(5)_R$ symmetry group
\beqa
\ga^\mu \ga^\nu +\ga^\nu\ga^\mu &=& -2\eta^{\mu\nu} \\
\ga^A_R\ga^B_R + \ga^B_R\ga^A_R &=& 2\de^{AB},
\eeqa
with $\mu=0,...,4$ and $\eta^{\mu\nu} = \textrm{diag}(-,+,+,+,+)$ and $A=1,...,5$.

\subsection{Outline}
\seclab{outline}
We want to show that for the scattering processes just described, all $S$ matrix elements are related by supersymmetry to one unknown function. In order to do that we will in section \secref{dof} pursue a rather lengthy discussion of the massless and massive vector multiplets. Pictorially, we will arrange the states of the two multiplets in two different diamonds, see Figure \figref{diamonds}
\begin{figure}
\centering
\includegraphics[width=6cm]{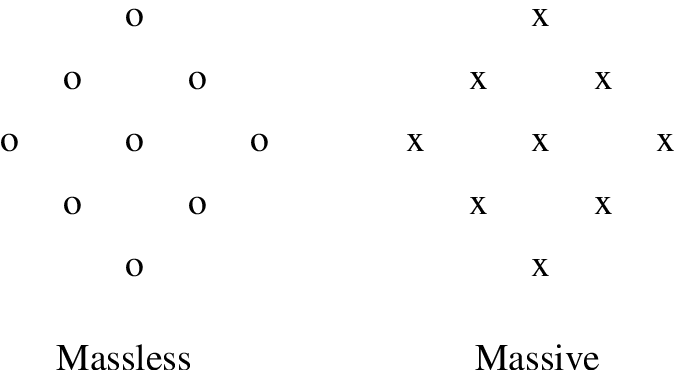}
\caption{The degrees of freedom of the massless and massive vector multiplets may be arranged in these diamond forms, see section \secref{dof}.}
\figlab{diamonds}
\end{figure}
where each ring and cross stands for a specific massless and massive polarization. We will argue that the supercharges can move around in these diamonds diagonally, \ie
\beq
\begin{array}{cc}
Q({\sss \nwarrow}) \nwarrow &  \nearrow Q({\sss \nearrow})\vspace{3mm} \cr 
Q({\sss \swarrow}) \swarrow & \searrow Q({\sss \searrow})
\end{array}
\eeq
This means that acting with $Q({\sss \nwarrow})$ on \eg the bottom state in a diamond gives the state above it to the left, etc. We will have one set of supercharges acting on the massless states and another set of supercharges acting on the massive states. In section \secref{2part} we will form two-particle states by taking the tensor product of the states in these diamonds. We then show how the supercharges act on the two-particle states. In doing this, we will automatically obtain the relations between the various $S$ matrix elements. We would like to stress that thinking of the supercharges and polarizations in this pictorial way will make the reading of the rather detailed section \secref{2part} much easier.

In section \secref{2,0} we relate our results to the analogous scattering processes in (2,0) theory. Most importantly, we calculate the low-energy result of the unknown function and compare it to its counterpart in the (2,0) theory. As already mentioned, they are equal.

\section{The degrees of freedom}
\seclab{dof}
\subsection{The massless vector multiplet}
\seclab{massless vm}
Let us now consider a massless vector multiplet particle with spatial momentum $\pbf$ and energy $\om$, such that $p^\mu = (\om;\pbf)$ and $p^2=0$. This breaks the Lorentz group to the little group $SO(3)_{\pbf}$ of spatial rotations that leave $\pbf$ invariant. For $\pbf$ in the $x^4$-direction, the supersymmetry algebra becomes
\beq
\{ Q,Q^\dagger \} = \ga^\mu p_\mu  = -\om  \ga^0({\bf 1} - \ga^{04}).
\eeq
The eight supercharges that are projected out by the right hand side are unbroken and annihilate such a particle state, apparently these have eigenvalue $+1$ under $\ga^{04}$. The remaining eight supercharges, having eigenvalue $-1$ under $\ga^{04}$, are broken and they can be used to construct the massless vector multiplet. Let us write $Q_u$ for the unbroken charges and $Q_b$ for the broken ones. Under the symmetry group $SO(3)_{\pbf} \times SO(4)_R$ the states of the massless vector multiplet transform as
\beq
\eqnlab{massless pol}
(3;1) \oplus (2;4_s) \oplus (1;1) \oplus (1;4_v).
\eeq
These are the degrees of freedom of the gauge field, the spinor, the Higgs boson and the remaining scalars of super Yang-Mills theory in the Coulomb phase. We may write any such state as $\left|\pbf,s \right>$, with $s$ running over all the sixteen different polarizations in \eqnref{massless pol}. To construct a basis for the polarization label $s$, we want to express the polarizations above in a way that preserves the symmetries of the scattering problem. (Recall from the previous section that we give the $SO(4)_R$ representation with the $\hat{J}$-eigenvalue as a subscript.) The gauge field is then decomposed as $1_{+1}, 1_0,1_{-1}$. The spinor becomes $2_{+1/2},2_{-1/2},2'_{+1/2},2'_{-1/2}$, with 2 and 2' referring to the two different spinor representations of $SO(4)$. The Higgs boson is also a scalar with zero-eigenvalue under $\hat{J}$, so to distinguish it from one of the polarizations of the gauge field, we call it $\tilde{1}_0$. Also the supercharges have a natural description in this language, they are $Q_b(2_{\pm1/2}),Q_b(2'_{\pm1/2})$ and $Q_u(2_{\pm1/2}),Q_u(2'_{\pm1/2})$. It is now natural to arrange the particle polarizations in the following way
\beq
 \eqnlab{sbasis}
\begin{array}{ccccc}
 & & \left| {\bf p}, 1_{+1} \right> & & \cr
 &  \left| {\bf p}, 2^\prime_{+1/2} \right> & &  \left| {\bf p}, 2_{+1/2} \right> & \cr
  \frac{1}{\sqrt{2}} (\left| {\bf p}, \tilde{1}_0 \right> -  \left| {\bf p}, 1_0 \right> ) & &  \left| {\bf p}, 4_0 \right> & & \frac{1}{\sqrt{2}} (\left| {\bf p}, \tilde{1}_0 \right> +  \left| {\bf p}, 1_0 \right>) \cr
 &  \left| {\bf p}, 2_{-1/2} \right> & &  \left| {\bf p}, 2^\prime_{-1/2}  \right> & \cr
 & &  \left| {\bf p}, 1_{-1} \right>
 \end{array} .
 \eeq
We then find that the broken supercharges, $Q_b$, act on the states in this diamond in the following manner
\beq
\eqnlab{susyarrows}
\begin{array}{cc}
Q_b(2_{+1/2}) \nwarrow &  \nearrow Q_b(2^\prime_{+1/2}) \vspace{3mm}\cr
Q_b(2^\prime_{-1/2}) \swarrow & \searrow Q_b(2_{-1/2}) 
\end{array}
\eeq
This means that acting with one of the $Q_b$ in the $2_{+1/2}$ represention on \eg the state $\left| \pbf,1_{-1}\right>$ gives the corresponding $\left| \pbf,2_{-1/2}\right>$ state, while it annihilates \eg the state $\left| \pbf,1_{+1}\right>$. By definition, the $Q_u$ charges annihilate any state $\left| \pbf,s\right>$. We need to make one extra comment about this, namely the supercharges are normalized such that $|Q| \sim \sqrt{\om}$, which implies that \eg $Q_b(2_{+1/2}) \left| \pbf,1_{-1}\right> = \sqrt{\om} \left| \pbf,2_{-1/2}\right>$.

We now turn to the question of how the supercharges (both $Q_b$ and $Q_u$) act on a more general state $\left| \pbf',s\right>$, with $\pbf'$ being any other momentum vector in the $x^3x^4$-plane. To answer this question, we must first give a proper definition of the state $\left| \pbf',s\right>$. Let us do that in the following way
\beq
\eqnlab{gen massless}
\left| \pbf',s\right> \equiv L(\pbf') \left| \pbf,s\right>,
\eeq
where $L(\pbf')$ is a standard Lorentz transformation of the vector $p^\mu$ to $p'^\mu$. We may take this to consist of first an appropriate Lorentz rotation in the $x^0x^4$-plane (boost) with rapidity $\eta$ (where $\exp{\eta} = |\pbf|/|\pbf'| \equiv \om/\om'$), followed by a spatial rotation in the plane spanned by $\pbf$ and $\pbf'$ by the angle $\theta$ between $\pbf$ and $\pbf'$. It then follows that $Q$ will have the same action on the state $\left| \pbf',s\right>$ as $L^{-1}(\pbf')QL(\pbf')$ has on the state $\left| \pbf,s\right>$. This is easily demonstrated as
\beq
\eqnlab{Qp'}
Q\left| \pbf',s\right> \equiv Q L(\pbf') \left| \pbf,s\right> = L(\pbf') \left(L^{-1}(\pbf') Q L(\pbf') \left| \pbf,s\right>\right),
\eeq
where now $L^{-1}(\pbf')QL(\pbf')$ acts on $\left| \pbf,s\right>$ (changing only the polarization) and then $L(\pbf')$ acts on the resulting state (changing only $\pbf$ to $\pbf'$). We will only be interested in situations where $\pbf'$ lies in the $x^3x^4$-plane and $|\pbf'| < |\pbf|$. In those cases we get
\beq
\eqnlab{LQL}
L^{-1}(\pbf') Q L(\pbf') = \left( {\bf 1} \cosh{\frac{\eta}{2}} + 2J^{04} \sinh{\frac{\eta}{2}} \right) \left( {\bf 1} \cos{\frac{\theta}{2}} + 2i J^{34} \sin{\frac{\theta}{2}} \right) Q,
\eeq
where $J^{04} \equiv \ga^{04}/2$ and $J^{34} \equiv i \ga^{34}/2$ are rotation generators in the spinor representation. Since $J^{04}J^{34} = - J^{34}J^{04}$, it follows that $Q_u$ will not annihilate a state $\left| \pbf',s\right>$ in general. Another crucial observation is that both $J^{04}$ and $J^{34}$ commute with $J^{12}$, which implies that the supercharges (both $Q_b$ and $Q_u$) in principle still act according to \eqnref{susyarrows} also on the states $\left| \pbf',s\right>$ when put in the diamond form of \eqnref{sbasis}. However, the action is now accompanied by some function of $\eta$ and $\theta$. To derive these functions, we consider first the case when a broken supercharge $Q_b$ acts on $\left| \pbf',s\right>$. According to \Eqsref{Qp'} and \eqnref{LQL} we then get
\begin{multline}
Q_b \left| \pbf',s\right> =\\
= L(\pbf') \left( {\bf 1} \cosh{\frac{\eta}{2}} + 2J^{04} \sinh{\frac{\eta}{2}} \right) \left( {\bf 1} \cos{\frac{\theta}{2}} + 2i J^{34} \sin{\frac{\theta}{2}} \right) Q_b  \left| \pbf,s\right> \\
= L(\pbf') \exp{\frac{-\eta}{2}} \cos{\frac{\theta}{2}} Q_b  \left| \pbf,s\right>,
\end{multline}
where the action of $Q_b$ on  $\left| \pbf,s\right>$ was described above. (If for example $Q_b \left| \pbf,s\right> = \sqrt{\om} \left| \pbf,\tilde{s}\right>$ with $\tilde{s}$ being some other polarization then $s$, then the right hand side becomes $\exp{\frac{-\eta}{2}} \cos{\frac{\theta}{2}} \sqrt{\om} \left| \pbf',\tilde{s}\right>$.) When deriving the analogous expression for an unbroken supercharge, we have to take its eigenvalue under $\hat{J}$ (which generates rotations in the $x^1x^2$-plane) into consideration:
\begin{multline}
\eqnlab{qup}
Q_u(2_{\pm1/2}) \left| \pbf',s\right> = \\
= L(\pbf') \left( {\bf 1} \cosh{\frac{\eta}{2}} + 2J^{04} \sinh{\frac{\eta}{2}} \right) \left( {\bf 1} \cos{\frac{\theta}{2}} + 2i J^{34} \sin{\frac{\theta}{2}} \right) Q_u(2_{\pm1/2})  \left| \pbf,s\right> \\
= \mp L(\pbf') \exp{\frac{-\eta}{2}} i\sin{\frac{\theta}{2}} \ga^0 Q_u(2_{\pm1/2}) \left| \pbf,s\right>.
\end{multline}
This relation looks exactly the same for the generators $Q_u(2'_{\pm1/2})$. By definition, $Q_u$ has eigenvalue $+1$ under $\ga^{04}$, and $Q_b$ eigenvalue $-1$. In the last line above, the operator $\ga^0 Q_u$ has eigenvalue $-1$ under $\ga^{04}$. Hence, given $Q_u$ in a specific representation ($2_{\pm 1/2}$ or $2'_{\pm 1/2}$), the new operator $\ga^0 Q_u$ must be proportional to the corresponding $Q_b$ in the same representation. From the superalgebra it follows that the two can only differ by a conventional sign, and we may choose $\ga^0 Q_u(2_{\pm1/2}) \equiv Q_b(2_{\pm1/2})$ and $\ga^0 Q_u(2'_{\pm1/2}) \equiv Q_b(2'_{\pm1/2})$.

For future reference, we end this section by summarizing the results just found:
\beqa
\eqnlab{hej}
Q_b(2_{\pm1/2}) \left| \pbf',s\right> &=& L(\pbf') \exp{\frac{-\eta}{2}} \cos{\frac{\theta}{2}} Q_b(2_{\pm1/2}) \left| \pbf,s\right> \\
Q_b(2'_{\pm1/2}) \left| \pbf',s\right> &=& L(\pbf') \exp{\frac{-\eta}{2}} \cos{\frac{\theta}{2}} Q_b(2'_{\pm1/2}) \left| \pbf,s\right> \\
\eqnlab{hopp}
Q_u(2_{\pm1/2}) \left| \pbf',s\right> &=& \mp L(\pbf')  \exp{\frac{-\eta}{2}} i\sin{\frac{\theta}{2}} Q_b(2_{\pm1/2}) \left| \pbf,s\right> \\
\eqnlab{hej sist}
Q_u(2'_{\pm1/2}) \left| \pbf',s\right> &=& \mp L(\pbf')  \exp{\frac{-\eta}{2}} i\sin{\frac{\theta}{2}} Q_b(2'_{\pm1/2}) \left| \pbf,s\right>,
\eeqa
where the supercharges $Q_b(2_{\pm1/2}) $ and $Q_b(2'_{\pm1/2}) $ act on the states $\left| \pbf,s\right>$ as described in \eqnref{sbasis} and \eqnref{susyarrows}. These last four equations are the main results of this section and we will put them to use in section \secref{2part}.

\subsection{The massive vector multiplet}
\seclab{massive vm}
We now go through the same procedure as in the previous subsection, but this time for the BPS saturated massive vector multiplet. The supersymmetry algebra in five dimensions includes a central charge $Z$ which is proportional to the vacuuum expectation value of the scalar fields of the super Yang-Mills theory at spatial infinity. Hence, it is a singlet under the Lorentz group, but breaks the $R$-symmetry group to $SO(4)_R$ and the gauge group to $U(1)$. This term gives rise to the Higgs mechanism, and from a six-dimensional perspective it corresponds to a string of tension $T$ winding the circle of compactification of radius $R$. Let us then consider a massive vector multiplet particle in its rest-frame,  \ie $k^\mu = (m;{\bf 0})$, and a central charge $Z=2\pi R T \ga^5_R$. Apparently, this setup also breaks the Lorentz group to the $SO(4)_\kbf$ little group of rotations that leave $\kbf$ invariant. The supersymmetry algebra becomes
\beq
\{ \tilde{Q},\tilde{Q}^\dagger \} = -m\ga^0 - 2\pi R T \ga^5_R.
\eeq
To find the short representation we choose $m = 2\pi RT$ and thus find that the eight supercharges which are annihilated by ${\bf 1} +\ga^0 \ga^5_R$ are left unbroken by such a configuration. The other eight supercharges are broken, and they may be used to construct the short massive vector multiplet representation. Before proceeding however, let us dwell a little on these supercharges: First of all we note that $\ga^0=\ga^{1234}$, so that spinors with positive (negative) eigenvalue under $\ga^0$ can be said to be in the (anti-) chiral representation of the $SO(4)_\kbf$ little group. Furthermore, $\ga^5_R = \ga^{1234}_R$ so that $SO(4)_R$-spinors with positive (negative) eigenvalue under $\ga^5_R$ can be said to be in the (anti-) chiral representation of the $R$-symmetry group. Let us always denote a chiral spinor by 2 and an anti-chiral spinor by 2'.\footnote{When talking about chiral and anti-chiral supercharges in the remainder of this paper, we will always refer to the chirality under the $SO(4)_\kbf$ little group.} It then follows that under the symmetry group $SO(4)_\kbf\times SO(4)_R$, the unbroken supercharges, $\tilde{Q}_u$, are
\beq
\eqnlab{mass un}
(2';2) \oplus (2;2')
\eeq
and the broken ones, $\tilde{Q}_b$, are
\beq
(2;2) \oplus (2';2').
\eeq
Notice that we may write these four sets of charges as $\tilde{Q}_u(2_{\pm1/2}),\tilde{Q}_u(2'_{\pm1/2})$ and $\tilde{Q}_b(2_{\pm1/2}),\tilde{Q}_b(2'_{\pm1/2})$ just as in the previous section. Under the preserved symmetry group $SO(4)_\kbf \times SO(4)_R$ the states of the massive vector multiplet transform as
\beq
(4_v;1) \oplus (2;2') \oplus (2';2) \oplus (1;4_v)
\eeq
corresponding to a vector field, chiral and anti-chiral spinor fields transforming as spinors also under $R$-symmetry and four scalar fields. Once again we may write any such state as $\left| \kbf , \si \right>$, with $\si$ running over the sixteen polarizations above. We now wish to construct a basis for the polarization label $\si$ that respects the symmetries of the scattering problem ($SO(2)_{\pbf,\kbf;\pbf';\kbf'} \times SO(4)_R$) and we do that in a manner analogous to the previous section. A natural way to describe the four degrees of freedom in the vector field would be to introduce four polarization vectors pointing in the $x^1$, $x^2$, $x^3$ and $x^4$-direction respectively. However, if we call these $\eps^1,...,\eps^4$ it is actually more convenient to combine them as $\eps^1 + i\eps^2$, $\eps^1 - i\eps^2$, $\eps^3 + i\eps^4$, $\eps^3 - i\eps^4$. These four combinations correspond to the following $\si$-polarizations: $1_{+1}, 1_{-1}, 1_0,1'_0$, where we put a prime on the last polarization to distinguish it from the third one. The first of the spinors is written $2'_{+1/2},2'_{-1/2}$, whereas the second spinor becomes $2_{+1/2},2_{-1/2}$. Finally, the scalars are $4_0$. Let us now arrange these states in the following way
\beq
 \eqnlab{sigmabasis}
\begin{array}{ccccc}
 & & \left| \kbf, 1_{+1} \right> & & \cr
 &  \left| {\bf k}, 2^\prime_{+1/2} \right> & &  \left| {\bf k}, 2_{+1/2} \right> & \cr
  \left| {\bf k}, 1'_0 \right> & &  \left| {\bf k}, 4_0 \right> & & \left| {\bf k},1_0 \right> \cr
 &  \left| {\bf k}, 2_{-1/2} \right> & &  \left| {\bf k}, 2^\prime_{-1/2}  \right> & \cr
 & &  \left| {\bf k}, 1_{-1} \right>
 \end{array} .
 \eeq
We then find that the broken supercharges act on the states in this diamond in the following manner
\beq
\eqnlab{susyarrows2}
\begin{array}{cc}
\tilde{Q}_b(2_{+1/2}) \nwarrow &  \nearrow \tilde{Q}_b(2^\prime_{+1/2}) \vspace{3mm}\cr
\tilde{Q}_b(2^\prime_{-1/2}) \swarrow & \searrow \tilde{Q}_b(2_{-1/2}) 
\end{array}
\eeq
while the unbroken supercharges annihilate any state in the diamond. Recall from the supersymmetry algebra that $|\tilde{Q}| \sim \sqrt{m}$, which implies \eg that $\tilde{Q}(2_{+1/2}) \left| \kbf , 1_{-1} \right> = \sqrt{m} \left| \kbf , 2_{-1/2} \right>$. Notice the analogy with the massless case. 

Continuing on these lines, we now want to examine how the supercharges act on a more general state $\left| \kbf' , \si \right>$, with $\kbf'$ being a spatial momentum vector in the $x^3x^4$-plane, however $k'_\mu k'^\mu = -m^2$. We define this general state in the following way
\beq
\eqnlab{gen massive}
\left| \kbf' , \si \right> \equiv \tilde{L}(\kbf')\left| \kbf , \si \right>,
\eeq
where $\tilde{L}(\kbf')$ is a standard rotation of the vector $k^\mu$ to $k'^\mu$. Just as in the previous section, we take this to consist of first a boost in the $x^0x^4$-plane with rapidity $\eta'$ ($\sinh{\eta'}=v\ga(v)$ and $\cosh{\eta'} = \ga(v)$, with $v$ being the speed of the particle and $\ga(v)$ the relativistic gamma-factor), followed by a spatial rotation in the $x^3x^4$-plane by the angle $\phi$ between $\hat{x}^4$ and $\kbf'$. It then follows that a supercharge $\tilde{Q}$ will have the same action on the state $\left| \kbf' , \si \right>$ as $\tilde{L}^{-1}(\kbf') \tilde{Q} \tilde{L}(\kbf')$ has on the state $\left| \kbf , \si \right>$. We will only be interested in cases where the $x^3$-component of $\kbf'$ has the opposite sign compared to that of $\pbf'$, because of conservation of momentum. It is also quite trivially true that $|\kbf'| > |\kbf|$. In these cases we get
\beq
\tilde{L}^{-1}(\kbf') \tilde{Q} \tilde{L}(\kbf') = \left( {\bf 1} \cosh{\frac{\eta'}{2}} - 2J^{04} \sinh{\frac{\eta'}{2}} \right) \left( {\bf 1} \cos{\frac{\phi}{2}} - 2i J^{34} \sin{\frac{\phi}{2}} \right) \tilde{Q}.
\eeq
We find that the unbroken charges, having opposite chirality under the little group and the $R$-symmetry group, will not in general annihilate a state $\left| \kbf' , \si \right>$, since both $J^{04}$ and the combination $J^{04}J^{34}$ change the chirality under the little group. We also note that the supercharges (both broken and unbroken) act according to \Eqnref{susyarrows2} also on the states $\left| \kbf' , \si \right>$, however the action is accompanied by some function of $\eta'$ and $\phi$. To derive these functions, consider first the broken generators:
\beqa
\eqnlab{ooops}
\tilde{Q}_b(2_{\pm1/2}) \left| \kbf' , \si \right> &=& L(\kbf') \cosh{\frac{\eta'}{2}} \exp{\frac{\pm i\phi}{2}} \tilde{Q}_b(2_{\pm1/2}) \left| \kbf, \si \right> \\
\tilde{Q}_b(2'_{\pm1/2}) \left| \kbf' , \si \right> &=& L(\kbf') \cosh{\frac{\eta'}{2}} \exp{\frac{\mp i\phi}{2}} \tilde{Q}_b(2'_{\pm1/2}) \left| \kbf , \si \right>.
\eeqa
For the unbroken generators we find that
\beqa
\eqnlab{hejsan}
\tilde{Q}_u(2_{\pm1/2}) \left| \kbf' , \si \right> &=& L(\kbf') \sinh{\frac{\eta'}{2}} \exp{\frac{\pm i\phi}{2}} \tilde{Q}_b(2_{\pm1/2}) \left| \kbf , \si \right> \\
\eqnlab{ooops sist}
\tilde{Q}_u(2'_{\pm1/2}) \left| \kbf' , \si \right> &=& -L(\kbf') \sinh{\frac{\eta'}{2}} \exp{\frac{\mp i\phi}{2}} \tilde{Q}_b(2'_{\pm1/2}) \left| \kbf , \si \right>,
\eeqa
where we have applied the following two definitions: $\ga^4 \tilde{Q}_u(2_{\pm1/2}) \equiv  \tilde{Q}_b(2_{\pm1/2})$ and $\ga^4 \tilde{Q}_u(2'_{\pm1/2}) \equiv  \tilde{Q}_b(2'_{\pm1/2})$, which is analogous to what we did below \Eqnref{qup} in the previous subsection.

\section{Two-particle states and scattering}
\seclab{2part}
In the scattering processes of our interest, both the incoming and outgoing states consist of two particles; one massless and one massive. This far, we have reviewed how the supercharges $Q_b,Q_u$ act on the massless vector multiplet and how the supercharges $\tilde{Q}_b,\tilde{Q}_u$ act on the massive vector multiplet. We now want to form two-particle states and examine how supersymmetry acts on these. In doing this, we will automatically obtain the symmetry relations between the $S$ matrix elements.

We form a general two-particle state with momenta in the $x^3x^4$-plane in the following way
\beq
\left| \pbf' , s \right> \otimes \left| \kbf' , \si \right>
\eeq
with the two factors defined as in \Eqsref{gen massless} and \eqnref{gen massive}.  An infinitesimal supersymmetry transformation acts on such a state in following way
\beq
\Q \left( \left| \pbf' , s \right> \otimes \left| \kbf' , \si \right> \right) = Q \left| \pbf' , s \right> \otimes \left| \kbf' , \si \right> + \left| \pbf' , s \right> \otimes \tilde{Q} \left| \kbf' , \si \right>
\eeq
with the two terms in the right hand side behaving as was described in the previous section. Another way to write this is: $\Q = Q \otimes \tilde{{\bf 1}} + {\bf 1} \otimes \tilde{Q}$.

Now comes the key observation of this paper: Because rotations in the $x^1x^2$-plane is a symmetry of the scattering processes of our concern, the eigenvalue under $\hat{J}$ of the in-state and the out-state must be the same. Let us then start our analysis by considering an in-state of the following form
\beq
\left| \pbf , 1_{-1} \right> \otimes \left| \kbf , 1_{-1} \right>
\eeq
which has eigenvalue $-2$ under $\hat{J}$. It is clear that this is the only two-particle state with this eigenvalue, hence the out-state must be of the same form, but with the momenta exchanged for $\pbf'$ and $\kbf'$ lying in the $x^3x^4$-plane. This implies that
\beq
\eqnlab{allmighty}
\Big( \left< \pbf', s \right| \otimes \left< \kbf', \si \right| \Big) \Big( \left| \pbf, 1_{-1} \right> \otimes \left| \kbf, 1_{-1} \right> \Big) = \de_{s, 1_{-1}} \de_{\si,1_{-1}} S(\theta,\om,m) \de^{(5)}(p+k-p'-k')
\eeq
with the function $S(\theta,\om,m) \equiv S_{1_{-1},1_{-1};1_{-1},1_{-1}}$ being the specific entry in the $S$ matrix for the scattering of two particles with polarization $1_{-1}$ into two particles with the same polarization, traveling in the $x^3x^4$-plane. (Note that by conservation of five-momentum $S$ is a function only of three variables, which we choose to be $\theta$, $\om \equiv |\pbf|$ and $m$.) Our claim is that this function, $S(\theta,\om,m)$, is the only function that need to be determined from a specific measurement or scattering calculation. All other entries, $S_{s,\si;s',\si'}$, in the $S$ matrix concerning this type of scattering are related to this function by supersymmetry.

Let us demonstrate how to proceed to calculate any other $S$ matrix element. An instructive example is to take $\left| \pbf , 2_{-1/2} \right> \otimes \left| \kbf , 2_{-1/2} \right>$ as our in-state. To relate this to the above, we start by acting with a supercharge on the state $\left| \pbf , 1_{-1} \right> \otimes \left| \kbf , 1_{-1} \right>$ in such a way that the massless particle polarization is changed from $1_{-1}$ to $2_{-1/2}$, but with the massive particle polarization unaltered. We are then to take $\Q$ to be of the type $2_{+1/2}$ and since it is to annihilate the massive particle it must be anti-chiral under the little group, see \eqnref{mass un}. We introduce the notation $\Q_{\tilde{u}}(2_{+1/2})$ for this supercharge and find
\begin{multline}
\Q_{\tilde{u}}(2_{+1/2}) \left( \left| \pbf , 1_{-1} \right> \otimes \left| \kbf , 1_{-1} \right> \right) = \\
= Q \left| \pbf , 1_{-1} \right> \otimes \left| \kbf , 1_{-1} \right> + \left| \pbf , 1_{-1} \right> \otimes \tilde{Q}_u \left| \kbf , 1_{-1} \right> \\
= Q \left| \pbf , 1_{-1} \right> \otimes \left| \kbf , 1_{-1} \right>.
\end{multline}
To evaluate the last line above, we must find out how an anti-chiral supercharge in the $2_{+1/2}$ representation acts on the state $\left| \pbf , 1_{-1} \right>$. Recall from section \secref{massless vm} that a supercharge with positive eigenvalue under $\ga^{04}$ annihilates such a state, whereas one with negative eigenvalue changes the polarization to $2_{-1/2}$. Furthermore, we introduced the definition $Q_u \equiv \ga^0 Q_b$, from which it follows that we can write an anti-chiral supercharge as $Q = (Q_b - Q_u)/\sqrt{2}$ (and a chiral one as $Q=(Q_b+Q_u)/\sqrt{2}$). Hence, we find
\beq
Q(2_{+1/2}) \left| \pbf , 1_{-1} \right> = \frac{1}{\sqrt{2}} Q_b(2_{+1/2}) \left| \pbf , 1_{-1} \right> = \sqrt{\frac{\om}{2}} \left| \pbf , 2_{-1/2} \right>.
\eeq
We thus conclude that
\beq
\Q_{\tilde{u}}(2_{+1/2}) \left( \left| \pbf , 1_{-1} \right> \otimes \left| \kbf , 1_{-1} \right> \right) =\sqrt{\frac{\om}{2}}  \left| \pbf , 2_{-1/2} \right> \otimes \left| \kbf , 1_{-1} \right>.
\eeq

We are now to act with another supercharge on the state $\left| \pbf , 2_{-1/2} \right> \otimes \left| \kbf , 1_{-1} \right>$ such that the massive particle polarization is changed from $1_{-1}$ to $2_{-1/2}$ but with the massless particle polarization unchanged. This supercharge must also be in the $2_{+1/2}$ representation but instead of having a specific chirality, it should have positive eigenvalue under $\ga^{04}$. We call this supercharge $\Q_{u}(2_{+1/2})$ and find
\beq
\Q_{u}(2_{+1/2}) \left( \left| \pbf , 2_{-1/2} \right> \otimes \left| \kbf , 1_{-1} \right> \right) = \left| \pbf , 2_{-1/2} \right> \otimes \tilde{Q} \left| \kbf , 1_{-1} \right>.
\eeq
Recalling that $\tilde{Q}_u \equiv -\ga^4 \tilde{Q}_b$ it follows that we can write a supercharge with positive eigenvalue under $\ga^{04}$ as $\tilde{Q} = (\tilde{Q}_b + \tilde{Q}_u)/\sqrt{2}$ (and one with negative eigenvalue as $(\tilde{Q}_b - \tilde{Q}_u)/\sqrt{2}$). Hence, in the expression above we can excange $\tilde{Q}$ for $\tilde{Q}_b/\sqrt{2}$ and obtain
\beq
\tilde{Q} \left| \kbf , 1_{-1} \right> = \sqrt{\frac{m}{2}} \left| \kbf , 2_{-1/2} \right> .
\eeq

We have now shown that
\beq
\eqnlab{qun}
\Q_{\tilde{u}}(2_{+1/2}) \Q_{u}(2_{+1/2}) \left( \left| \pbf , 1_{-1} \right> \otimes \left| \kbf , 1_{-1} \right> \right) = \frac{\sqrt{\om m}}{2} \left| \pbf , 2_{-1/2} \right> \otimes \left| \kbf , 2_{-1/2} \right>.
\eeq
This can be used to calculate the following $S$ matrix elements:
\beq
\eqnlab{ex1}
\Big( \left< \pbf', s \right| \otimes \left< \kbf', \si \right| \Big) \Big( \left| \pbf, 2_{-1/2} \right> \otimes \left| \kbf, 2_{-1/2} \right> \Big) \equiv S_{2_{-1/2},2_{-1/2};s,\si} \de^{(5)}(p+k-p'-k').
\eeq
Let us choose $s=2_{-1/2}$ and $\si=2_{-1/2}$ and see what we get. By rewriting the kets according to \Eqnref{qun} and taking the Hermitean conjugate twice it follows that the relevant calculation becomes
\beq
\eqnlab{daggers}
\left( \Q^\dagger_{u}(2_{+1/2}) \Q^\dagger_{\tilde{u}}(2_{+1/2}) \left( \left| \pbf' , 2_{-1/2} \right> \otimes \left| \kbf' , 2_{-1/2} \right> \right) \right)^\dagger,
\eeq
where the Hermitean conjugate of the supercharges only changes the sign of their eigenvalue under $\hat{J}$. Hence, \eqnref{daggers} becomes (omitting the overall Hermitean conjugate)
\beq
\eqnlab{no daggs}
\left( Q_u \otimes \tilde{{\bf 1}} + {\bf 1} \otimes \frac{\tilde{Q}_b + \tilde{Q}_u}{\sqrt{2}} \right) \left( \frac{Q_b - Q_u}{\sqrt{2}} \otimes \tilde{\bf 1} + {\bf 1} \otimes \tilde{Q}_u \right) \left( \left| \pbf' , 2_{-1/2} \right> \otimes \left| \kbf' , 2_{-1/2} \right> \right) 
\eeq
with all supercharges in the $2_{-1/2}$ representation. First, we must act with the operators in the right parenthesis and by using \Eqsref{hej}, \eqnref{hopp} and \eqnref{hejsan} this gives
\beq
\sqrt{\frac{\om}{2}} \exp{\frac{-\eta}{2}} \exp{\frac{-i\theta}{2}}  \left| \pbf' , 1_{-1} \right> \otimes \left| \kbf' , 2_{-1/2} \right> + \sqrt{m} \sinh{\frac{\eta'}{2}} \exp{\frac{-i\phi}{2}} \left| \pbf' , 2_{-1/2} \right> \otimes \left| \kbf' , 1_{-1} \right>
\eeq
Then acting with the operators in the left parenthesis in \eqnref{no daggs}, we obtain
\beq
\eqnlab{yo}
\frac{\sqrt{\om m}}{2} \exp{\frac{-\eta}{2}} \exp{\frac{-i\phi}{2}} \left( \exp{\frac{\eta'}{2}} \exp{\frac{-i\theta}{2}} + 2 i\sin{\frac{\theta}{2}} \sinh{\frac{\eta'}{2}} \right) \left| \pbf' , 1_{-1} \right> \otimes \left| \kbf' , 1_{-1} \right>
\eeq
A little bit of relativistic geometry gives that $\cosh{\frac{\eta'}{2}} = \sqrt{(\ga + 1)/2}$ and $\sinh{\frac{\eta'}{2}} = \sqrt{(\ga -1)/2}$. Putting the pieces together we find that
\beq
\eqnlab{hola}
S_{2_{-1/2},2_{-1/2};2_{-1/2},2_{-1/2}} = \sqrt{\frac{\om'}{2\om}} \exp{\frac{i\phi}{2}} \left( \sqrt{\ga+1} \exp{\frac{i\theta}{2}} + \sqrt{\ga - 1} \exp{\frac{-i\theta}{2}} \right) S(\theta,\om,m).
\eeq
We note that it is possible to express $\ga, \om'$ and $\phi$ as functions of $\theta, \om$ and $m$, but that would make the expression rather ugly.

For an arbitrary choice of $s$ and $\si$ in \eqnref{ex1}, the expression in \eqnref{yo} becomes a linear combination of maximally three different two-particle states. In order for the left hand side in \Eqnref{ex1} to become non-zero, one of these must be the state $\left| \pbf' , 1_{-1} \right> \otimes \left| \kbf' , 1_{-1} \right>$ (by virtue of \Eqnref{allmighty}). It is easily realized that only three different choices of $s$ and $\si$ does that, namely the one we have studied $\left\{s=2_{-1/2},\si=2_{-1/2} \right\}$ together with $\left\{s=(\tilde{1}_0-1_0)/\sqrt{2},\si=1_{-1} \right\}$ and $\left\{s=1_{-1},\si=1'_0 \right\}$. To calculate the $S$ matrix elements for these two latter choices of $s$ and $\si$ one should follow exactly the procedure in the example we have just done.

Having read this far, it should be obvious that we can relate any $S$ matrix element to the function $S(\theta,\om,m)$ by using this method. We thus conclude that supersymmetry relates all possible scattering processes to one unknown function, which will be discussed in the next section.

\section{Relation to (2,0) theory}
\seclab{2,0}
In this section, we will relate our results to the analogous scattering problem in the six-dimensional (2,0) theory, where a massless particle is scattered off an infinitely long, tensile string. We will also determine the so far unknown function $S(\theta,\om,m)$ and compare it to the corresponding function in (2,0) theory. To make a quantitative comparison, the momentum of the particle in six dimensions must make a right angle with the string. Effectively, this turns the six-dimensional problem into an infinite set of equivalent five-dimensional problems, one for each point on the string. Furthermore, we note that the combination of infinite length and finite tension, $T$, renders the string infinitely massive. We therefore expect to find some similar features in the two theories in the limit where the mass of the target particle in five dimensions is very large.

\subsection{The symmetry relations}
The way to compare the symmetry relations in five dimensions to the (2,0) theory results obtained in \cite{Flink:2005}, is to take the low-energy limit. In this limit we can regard the massive particle as being at rest also in the out-state. We can therefore expand the relations \eqnref{hej}-\eqnref{hej sist} and \eqnref{ooops}-\eqnref{ooops sist} implied by supersymmetry to lowest order in the parameter $\om/m$ by setting $\eta'=\eta=\phi=0$. (Recall that $\om$ is the energy of the incoming massless particle, and $m$ is the mass of the massive particle.) We may then use the supercharges $\Q_u$ (that do not affect the polarization of a massless particle traveling in the $x^4$-direction) to show that the polarization of the massive particle does not change to lowest order in $\om/m$. We can also use these supercharges to show that the scattering amplitudes are independent of the massive particle polarization (as long as it does not change). Let us illustrate the first claim with an example: Consider a scattering process with $\left| \pbf,1_{-1} \right> \otimes \left| \kbf,2_{-1/2} \right>$ as in-state, and $\left| \pbf',2_{-1/2} \right> \otimes \left| \kbf,1_{-1} \right>$ as out-state. (Of course, the massive particle is not truly at rest in the out-state, since that would violate the conservation of momentum.) The corresponding $S$ matrix element is:
\begin{multline}
\left(\left< \pbf' , 2_{-1/2} \right| \otimes \left< \kbf ,1_{-1}\right|\right) \left( \left| \pbf,1_{-1} \right> \otimes \left| \kbf,2_{-1/2} \right> \right) = \\
= \left(\left< \pbf' , 2_{-1/2} \right| \otimes \left< \kbf ,1_{-1}\right|\right) \sqrt{\frac{2}{m}}\Q_u(2_{+1/2})\left( \left| \pbf,1_{-1} \right> \otimes \left| \kbf,1_{-1} \right> \right) \\
= \left( \left(\left< \pbf , 1_{-1} \right| \otimes \left< \kbf ,1_{-1}\right|\right) \sqrt{\frac{2}{m}}\Q_u(2_{-1/2})\left( \left| \pbf',2_{-1/2} \right> \otimes \left| \kbf,1_{-1} \right> \right) \right)^\dagger \\
= \left( \sqrt{\frac{2\om}{m}} i\sin{\frac{\theta}{2}} \left(\left< \pbf , 1_{-1} \right| \otimes \left< \kbf ,1_{-1}\right|\right) \left( \left| \pbf',1_{-1} \right> \otimes \left| \kbf,1_{-1} \right> \right) \right)^\dagger \\
= -\sqrt{\frac{2\om}{m}} i\sin{\frac{\theta}{2}} S(\theta,\om,m) \de(|\pbf|-|\pbf'|),
\end{multline}
where we have used that $\Q_u^\dagger(2_{+1/2}) = \Q_u(2_{-1/2})$ and \Eqnref{hopp}. Proceeding like this, one finds that all amplitudes in which the massive particle changes its polarization are either identically zero or includes at least an extra factor of $\sqrt{\om/m}$ times $S(\theta,\om,m)$.  Hence, to lowest order in $\om/m$, these amplitudes are all zero. It now follows for free that neither the massless particle polarization is changed in the low-energy limit.

Let us then take the massive particle to have polarization $1_{-1}$ and investigate how the $S$ matrix depends on the massless particle polarization. We are to make use of the supercharges $\Q_{\tilde{u}}$ that do not affect the polarization of a massive particle at rest. (Since the massive particle can be treated as being at rest in both the in- and out-state, these supercharges will not affect the massive particle polarization at all.)  It is then straight forward to derive the following relations: 
\beq
 \eqnlab{scat}
\begin{array}{ccccc}
 & & 1 & & \cr
 & e^{\frac{i\theta}{2}}  & & e^{\frac{-i\theta}{2}}  & \cr
 e^{i\theta} & & 1 & & e^{-i\theta} \cr
 &  e^{\frac{i\theta}{2}} & &  e^{\frac{-i\theta}{2}} & \cr
 & &  1
 \end{array} 
 \eeq
The results in \eqnref{scat} are to be interpreted as follows: The $S$ matrix element for scattering of a massless particle with polarization $2_{-1/2}$ is $e^{\frac{i\theta}{2}} S(\theta,\om,m)$ etc. These results are in perfect agreement with the analogous scattering processes in (2,0) theory \cite{Flink:2005}.\footnote{The angular dependence looks a little bit different in \cite{Flink:2005}. However, the difference has only to do with the choice of basis for the degrees of freedom. Here we have the $x^4$-axis as our reference direction, while the $x^5$-axis was used in the cited paper.}

Basically, the agreement is due of the preserved symmetries being the same in the two problems, and the degrees of freedom being more or less one to one. Most importantly, the supersymmetry algebra in the super Yang-Mills theory can be derived from the superalgebra in the (2,0) theory. The difference between the two theories is that the string mass is truly infinite, while the mass of the massive particles in super Yang-Mills theory is only much much larger than the energy of the massless particles. Indeed, proceeding beyond leading order, the massive particle polarization can change in the five-dimensional processes, leaving us with the much more difficult analysis of sections \secref{dof} and \secref{2part}.

\subsection{The unknown function}
It is also very interesting to calculate the function $S(\theta,\om,m)$ to lowest order in $\om/m$ and compare it to the corresponding function $S_{(2,0)}(\theta,\om,T)$ in (2,0) theory (recall the relation $m=2\pi RT$ between the five-dimensional mass and the string tension). In \cite{Flink:2005} we found the surprisingly simple result
\beq
S_{(2,0)}(\theta,\om,T) \sim \frac{(2\pi)^{-3}}{\om T} +...
\eeq
where the dots indicate terms which are of higher order in $\om^2/T$. We notice that the result is actually independent of $\theta$ to lowest order. To determine $S(\om,\theta,m)$, we make use of the super Yang-Mills action
\beq
I = \int d^5x Tr \big\{ F_{\mu\nu} F^{\mu\nu} + D_{\mu} \phi_AD^\mu \phi_A + g^2\left[ \phi_A, \phi_B \right] \left[ \phi_A, \phi_B \right] + \bar{\psi}_i D \!\!\!\! \slash \psi^i + g\phi_A \bar{\psi}_i \psi^j (\ga_R^A)^i_{\ph{i}j} \big\}
\eeq
with all the fields in the adjoint representation of the $SU(2)$ gauge group. The fields are related to the fields in the six-dimensional tensor multiplet as being their zero-modes in a Kaluza-Klein reduction (multiplied by $\sqrt{2\pi R}$). The coupling constant is $g^2=2\pi R$. The index $i=1,...,4$ is an $SO(5)_R$ spinor index, while the rest of the notation needs no further explanation. To use this action for our purpose, we must break the gauge group to $U(1)$ by giving the scalar fields a certain vacuum expectation value, a convenient choice is $\left< \phi_A^a \right> = \de_A^5\de^a_3 \frac{m}{g}$ with $a=1,2,3$ being a gauge index. The vector field $A^a_\mu$ then decomposes into a massless gauge field $A_\mu^3$ and two oppositely charged massive vector fields $A_\mu^+ \equiv A_\mu^1+iA_\mu^2$ and $A_\mu^-\equiv A_\mu^1-iA_\mu^2$. We then expand the trace and keep only the few interaction terms relevant (to tree level) for the scattering of a massless vector field against a massive vector field:
\beq
I = \int d^5x \left(g^2 A_\mu^3 A^{3\mu} A_\nu^+ A^{-\nu} - \eps_{abc} g \pa_{[\mu} A_{\nu]}^a A^{b\mu} A^{c\nu} +...\right),
\eeq
where we have made no effort in getting the numerical coefficients right. The second term is a three-vertex which can be used to make two different Feynman diagrams to tree level (different channels), both involving a massive propagator. It turns out that in the special case where we take all four external particles to have polarization $1_{-1}$, both these two diagrams are zero. This can be understood by noticing that there is no propagator which can carry eigenvalue $-2$ under $\hat{J}$. Hence, to tree level, the whole contribution to this scattering amplitude comes from the first term, which is a four-vertex. Using the Feynman rules of \cite{Weinberg_1} we find that
\begin{multline}
S_{1_{-1},1_{-1};1_{-1},1_{-1}} \sim \\
\sim \frac{g^2(2\pi)^{-3}}{m\om} \de^{(5)}(p+k-p'-k') \eps_\mu^3(\pbf,1_{-1}) (\eps^{3\mu})^*(\pbf',1_{-1}) \eps_\nu^+(\kbf,1_{-1}) (\eps^{+\nu})^*(\kbf',1_{-1})  +...
\end{multline}
with the epsilons (one for each external leg) being the polarization vectors in Fourier space and the dots indicating higher order terms in $\om/m$. It is easily realized that
\beq
\eps_\mu^3(\pbf,1_{-1}) = \eps_\mu^3(\pbf',1_{-1}) = \eps_\mu^+(\kbf,1_{-1}) = \eps_\mu^+(\kbf',1_{-1}) = (0;1,i,0,0)/\sqrt{2}.
\eeq
Inserting this in the expression above and exchanging $g$ and $m$ for $R$ and $T$, we end up with
\beq
S(\theta,\om,T) \sim  \frac{(2\pi)^{-3}}{\om T}  +...
\eeq
which is the same result that we found for the corresponding function $S_{(2,0)}(\theta,\om,T)$ in the (2,0) theory. Before discussing this agreement, it is important to notice that it cannot persist to higher orders. Firstly, if we want to take loops into account, we would run into the problem of the super Yang-Mills theory being non-renormalizable. Furthermore, we have seen that the polarizations of the particles can change in super Yang-Mills theory already at tree-level. Hence, the theories show very different behaviors already to the next order in perturbation theory.

At first, it might seem surprising that the functions $S_{(2,0)}(\theta,\om,T)$ and $S(\theta,\om,m=2\pi R T)$ agree in the low-energy limit, since the calculations in five and six dimensions bear no resemblance at all. Furthermore, the compactification from six to five dimensions introduces new properties and difficulties. However, letting the incoming particle in the six-dimenisional problem make a right angle with the string, it follows from \cite{Flink:2005} that to lowest order in $\om^2/T$ the outgoing radiation must also consist of a single particle making a right angle with the string, and no string waves will be excited. This implies that we are not making use of the extra dimension in which the string is extended, so effectively we are left with a five-dimensional problem. In this five-dimensional hyperplane, the string looks like a particle. Identifying the ends of the string, being at infinity, would make no difference for the scattering problem. Then contracting the extra dimension (and thus also the string) to a finite length, has the important effect of making the string mass finite. However, to lowest order in $\om^2/T$ we can not measure such effects, since keeping only the lowest order terms we may neglect the movement of the massive particle/string in the out-state. We are thus left with the situation where a massless tensor multiplet particle is scattered of a massive particle in a five-dimensional subspace. The polarizations of these two constituents are in one to one correspondence with the massless and massive vector multiplets in the super Yang-Mills theory. We have thus arrived at the following two statements: 1) This compactified theory must give the same result for the function $S_{(2,0)}(\theta,\om,T)$ as the uncompactifed (2,0) theory, to lowest order in $\om^2/T$. 2) If (2,0) theory is to provide the UV-completion of  maximally supersymmetric Yang-Mills theory in five dimensions, the compactified theory we have arrived at must be described by the super Yang-Mills theory at low energies. Then, the functions $S_{(2,0)}(\theta,\om,T)$ and $S(\theta,\om,m=2\pi RT)$ must agree to lowest order in $\om^2/T$. We thus interpret our results as a confirmation of the belief that (2,0) theory should provide the UV-completion of five-dimensional super Yang-Mills theory.

\vspace{1cm}
\noindent
\textbf{Acknowledgments:}  I want to thank my supervisor M\aa ns Henningson and P\"ar Arvidsson for many enlightening discussions.

\newpage

\bibliographystyle{utphysmod3b}
\bibliography{paper}

\end{document}